# LEARNING LOCAL DISTORTION VISIBILITY FROM IMAGE QUALITY

*Navaneeth Kamballur Kottayil*[1], *Giuseppe Valenzise*[2], *Frederic Dufaux*[2] *and Irene Cheng*[1]

[1] MRC, University of Alberta, [2] CNRS, CentraleSupélec, Université Paris-Sud.

## ABSTRACT

Accurate prediction of local distortion visibility thresholds is critical in many image and video processing applications. Existing methods require an accurate modeling of the human visual system, and are derived through pshycophysical experiments with simple, artificial stimuli. These approaches, however, are difficult to generalize to natural images with complex types of distortion. In this paper, we explore a different perspective, and we investigate whether it is possible to learn local distortion visibility from image quality scores. We propose a convolutional neural network based optimization framework to infer local detection thresholds in a distorted image. Our model is trained on multiple quality datasets, and the results are correlated with empirical visibility thresholds collected on complex stimuli in a recent study. Our results are comparable to state-of-the-art mathematical models that were trained on phsycovisual data directly. This suggests that it is possible to predict psychophysical phenomena from visibility information embedded in image quality scores.

## 1. INTRODUCTION

Predicting the visibility of visual distortion is of paramount importance in a number of image processing applications, such as image compression [1, 2], watermarking [3] and quality assessment [4, 5]. Models of contrast detection and masking have been widely studied in the literature, e.g., [4, 6, 7]. These studies are based on an accurate modeling of the human visual system (HVS), including intra-ocular scattering, contrast sensitivity at different frequencies, luminance masking, intra- and inter-channel masking, etc.

Traditionally, measurements of distortion visibility are obtained through psychophysical experiments employing simple, artificial stimuli, such as Gabor patterns, sine-wave gratings, or wide-band noise [8, 6]. The simplicity and non-naturalness of the stimuli enable to describe them through well-defined features (e.g., frequency, brightness, etc.). These features are then used to derive mathematical models of distortion visibility. In order to understand how these observations generalize to the case of natural image masks, Alam *et al.* [9] have recently presented a dataset of local masking for natural scenes. Specifically, they use Gabor patterns as stimuli and natural images as masks to measure detection thresholds – the minimum magnitude of the stimulus making it distinguishable from the background mask – for small image patches. Later, this data has been used to train a convolutional neural network in order to predict the detection threshold of a patch from its input pixel values [10].

Conventional approaches to model distortion visibility strongly rely upon psychophysical experiments that are, in their nature, based on a simplification of real-world conditions. For example, models that describe visibility of sine-wave gratings might be enough to predict visibility in DCT-based image compression; however, they are probably failing in modeling different or multiple concurrent artifacts. Furthermore, local visibility is influenced by surrounding regions, and is ultimately linked to image semantics. It is evident that modeling all these complex factors *only* through psychophysical experiments is unfeasible.

In this work, we explore a different perspective. Instead of learning distortion visibility directly from psychophysical data, we propose to learn it *indirectly*, leveraging the large availability of alternative, yet related, data: subjectively annotated image quality assessment (IQA) datasets. Image quality scores provide higher level information about the visual appearance of a picture, compared to psychophysical measurements. At the same time, they bring information about the visibility of distortion. Indeed, a common assumption in image quality assessment is that the perceived quality is directly related to the visibility of the error signal [11, 12]. In other words, the per pixel error is weighted locally by the ensemble of perceptual phenomena, such as contrast sensitivity and several forms of masking, which discount its visibility to the human visual system.

Visual quality scores thus implicitly embed latent information on error visibility. In our recent work [13], we have proposed a deep convolutional neural network (CNN) architecture to disentangle the per pixel distortion and what we called the "perceptual resistance", in the context of no-reference quality estimation of high dyanamic range compressed pictures. Our results demonstrated that it is possible to effectively estimate these two terms over a broad range of qualities, starting from supra-threshold quality scores. In this paper, we investigate whether a similar approach can also be used to predict near-threshold visibility. To this end, we train our proposed system in a full-reference fashion, i.e., we assume that the error signal is known, as it is the case in most

IQA datasets. However, the inference step does not require the knowledge of the error, and can produce an estimate of local masking for any input image. Interestingly (and perhaps surprisingly), we find that perceptual scaling learned from image quality scores can predict the detection thresholds in [9] with similar accuracy as the CNN-based regressor in [10], although our model is learned on other datasets with different contents and several kinds of visual impairments. This makes the proposed approach potentially more general than previous work.

The rest of the paper is organized as follows. In Section 2, we describe our model to derive visibility threshold from image quality scores. In Section 3, we compare our predictions to state-of-the-art models on a local masking visibility dataset. Finally, we draw conclusions in Section 4.

## 2. PROPOSED MODEL

In this section we present our proposed model to estimate local distortion visibility thresholds. Specifically, we first discuss the assumptions of our model and provide a mathematical framework to split (supra-threshold) quality scores into per pixel error and a perceptual scaling term, which we show in Section 3 to be a good predictor of visibility thresholds in local masking. Afterwards, we describe how to implement this model using a deep convolutional neural network architecture.

### 2.1. Mathematical framework and assumptions

Let $I_R$ and $I_D$ be an original reference image and its distorted version, respectively, and $Q \in [0, 1]$ the quality score for $I_D$. Without loss of generality, we assume that $Q$ is given as a differential mean opinion score (DMOS). We assume that we have access to the local quality $q(i, j)$ of an image patch $I_D(i, j)$ of size $N \times N$ pixels, centered at location $(i, j)$. The pooling process linking the quality of individual patches to the overall quality $Q$ may depend on many factors, e.g., saliency [14]. Here, we assume that the local quality of an image block is the same as the global image quality score, similarly to the setting in [15]. While this is a strong assumption, which is often not met in practice, it has been proved to be accurate enough to predict image quality [15, 13].

In order to model local quality, we further assume that per pixel distortion in a patch is discounted by a perceptual weight, $T(i, j)$, that accounts for typical masking and visibility effects. Specifically, we measure pixel distortion through the average absolute error $E(i, j)$ of a patch, defined as:

$$E(i,j) = \frac{1}{N^2} \sum_{k=1}^{N^2} |I_D(i,j)_k - I_R(i,j)_k|, \quad (1)$$

where $k$ is the pixel index in the patch. We then approximate local quality as a function of the error and the perceptual

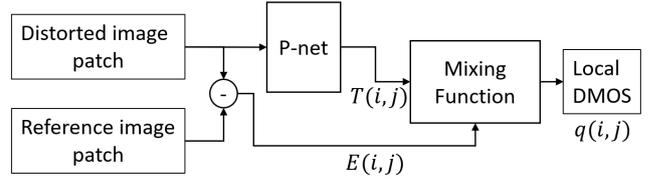

**Fig. 1**: Neural network architecture for extracting contrast detection thresholds from IQA databases.

weight, that is:

$$q(i,j) \approx 1 - \exp\left(-\left|\frac{\alpha \cdot E(i,j)}{T(i,j)}\right|^\beta\right), \quad (2)$$

where $\alpha$ and $\beta$ are two model parameters.

Notice that this formulation has been previously used to model the probability of detecting localized noise distortion, e.g., in [11], but we introduce an additional scaling factor $\alpha$. Eq. (2) is inspired by the common practice in vision science of expressing the magnitude of the error in multiples of the just-noticeable difference [2], although the relationship in this case is nonlinear. We therefore refer to $T(i, j)$ as the *local visibility threshold*, even if this is technically correct only for near-threshold distortion. We show in Section 3 that, for this latter case, $T$ does indeed model local visibility. Moreover, the thresholds we estimate are not in the same scale as those obtained through psychophysical experiments. We compensate for this by means of the parameter $\alpha$, which provides an additional degree of freedom in the optimization to push the predicted local quality as close as possible to ground truth. The value of the parameter $\beta$ is generally found by matching the results of psychophysical experiments. In practice, we found that the choice of $\beta$ does not significantly affect the performance of our model, and $\alpha$ alone provides already enough flexibility to minimize the loss function. Hence, we simply set $\beta = 1$ in Eq. (2).

### 2.2. Implementation

A scheme of the proposed model is shown in Figure 1. The input of our systems are overlapping patches of $32 \times 32$ pixels, similar to [15, 13]. The distortion, $E(i, j)$, is computed directly as in Eq. (1) during the training, where both $I_D$ and $I_R$ are available. The local visibility thresholds $T(i, j)$ are computed in a module that we name "P-net", while an estimate of local quality, $\hat{q}(i, j)$ is obtained by implementing Eq. (2) in the "Mixing function" block. Notice that this structure is required for training the P-net, as $T$ is considered a latent variable which depends implicitly on the observations of the input content and perceived quality. Instead, for inference the P-net is employed as a standalone block. Furthermore, we are generally interested in applying the learned P-net on the *original* pictures, rather than on the distorted ones. However, we found that training the P-net with noisy versions of the image was more effective, as this increases the variability of input data, leading to improved generalization capabilities.

Our model is trained to predict local quality $\hat{q}(i,j)$ using the ground-truth quality $q(i,j)$ as targets, by minimizing the following cost function:

$$J(i,j) = |q(i,j) - \hat{q}(i,j)|. \quad (3)$$

Notice that $\hat{q}(i,j)$ depends implicitly on the latent variables $T(i,j)$ through Eq. (2). Thus, when optimizing $J$, the visibility thresholds are adjusted in such a way to weigh the error coherently with the observed ground-truth quality.

For the architecture of P-net, we make use of a handcrafted layer that we named as augmented input layer [13]. In this layer, in addition to the luminance values of the $N \times N$ block, we compute the mean, variance and Mean Subtracted Contrast Normalized (MSCN) image [16]. The latter is defined as:

$$MSCN(x,y) = \frac{I(x,y) - \mu_M[I(x,y)]}{\sigma_M[I(x,y)] + \epsilon}, \quad (4)$$

where $(x,y)$ denotes the location of a pixel in the patch, $\mu_M[I(x,y)]$ is the mean and $\sigma_M[I(x,y)]$ the variance of the patch, computed by replacing every pixel $(x,y)$ with the mean and variance, respectively, over a local Gaussian window of size $M \leq N$ around $(x,y)$. The regularization term $\epsilon$ is set to 0.01. This is followed by convolutional layers with $32 \times 5 \times 5$ filters and a fully connected layer with 100 nodes. We use *relu* activation in all neurons. Dropout layers are used to prevent overfitting.

## 3. RESULTS AND ANALYSIS

In order to assess how well the estimated distortion visibility predicts ground-truth data from psychophysical experiments, we test the proposed model on the dataset of local masking thresholds in [9]. This dataset collects measured threshold values for 1080 image patches of size $85 \times 85$ pixels, extracted from the CSIQ dataset [17]. The detection thresholds are reported in terms of root-mean-squared (RMS) contrast and expressed in decibels (dB).

To test the proposed model, we train it on three different datasets: the CSIQ dataset [17], containing 855 images, 6 types of distortions; the TID 2013 dataset [18], with 3000 images and 25 types of distortion; and the LIVE dataset [19], featuring 779 images and 5 types of distortion. In general, the thresholds $T(i,j)$ found with our model do not lie on the same scale as those in [9]. In addition, the P-net can be trained on different IQA datasets, and the interpretation of DMOS in each dataset depends on the experiment carried out to collect the data [20]. Following a typical protocol in the evaluation of quality metrics [21], we compensate for this mismatch by linearizing the predictions with respect to psychophysical groundtruth through a monotonic third-order polynomial fitting before evaluating their statistical accuracy.

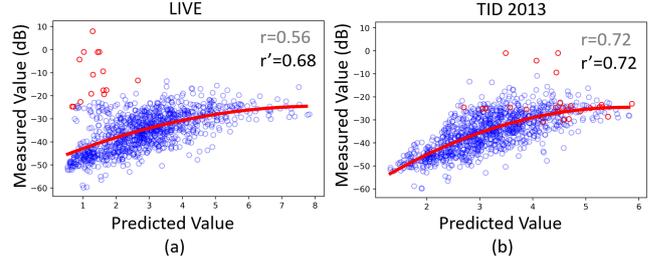

**Fig. 2**: Scatter plots of contrast detection thresholds derived using our method vs experimentally measured values. The system is trained on (a) LIVE dataset (b) TID 2013 dataset. The polynomial fitting line is show in red. Red points correspond to patches whose luminance is outside the range of the training datasets, see Figure 3. $r$ is the PLCC (after fitting) on the whole test set; $r'$ is the PLCC excluding the red points.

### 3.1. Performance

A comparison of the predicted and ground-truth local visibility thresholds is illustrated in the scatter plots of Figure 2, where our P-net has been trained on the LIVE and TID 2013 datasets, respectively. Each point in the scatter plot represents a $85 \times 85$ patch of [9]. Since our predictor can produce per pixel estimates of distortion visibility (using overlapping patches), we decimate the maps produced by the P-net to match the resolution of the ground truth, using a simple averaging filter (see Figures 4 and 5).

We can observe from Figure 2 that the predicted thresholds capture relatively well the overall trends of the measures obtained from psychophysical experiments, even if they have been obtained by training on very different data (IQA scores) and using different source contents. This indicates that learning visibility thresholds from generic IQA datasets is feasible and can generalize sufficiently well. Nevertheless, we notice in Figure 2 that in some cases the predicted thresholds deviate significantly from the measured ones. Especially for the LIVE dataset, this degrades the performance, measured by the Pearson linear correlation coefficient (PLCC) $r = 0.56$. To further investigate this, we analyze the distribution of the average luminance intensity of patches in the TID and LIVE datasets compared to CSIQ in Figure 3. Notice that there is only a negligible amount of patches for the LIVE dataset in the intensity range $[0, 10]$ and $[250, 255]$. This lack of data can affect the performance of prediction in this specific intensity range. We verify this by highlighting in red those patches of the test set having luminance outside the interval $[10, 250]$ in Figure 2. We observe that these points correspond indeed to the outliers in the scatter plot. By removing these few very dark or very bright patches (28 patches out of 1080), we ob-serve that the PLCC increases to $r' = 0.68$.

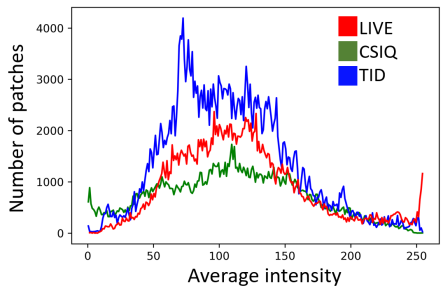

**Fig. 3**: Distribution of average intensity of $32 \times 32$ patches in LIVE, CSIQ and TID datasets. The distributions of the three datasets overlap only in the intensity range $[10, 250]$.

| Method | Training Data | RMSE |
|---|---|---|
| **Watson et al.-KMF [6]** | Pshyophysical visibility experiments | **5.713** |
| Watson et al.-JYS [6] | | 6.521 |
| Teo & Heeger [22] | | 6.861 |
| Chandler et al. [1] | | 6.879 |
| **Optimized GC [10]** | | **5.192** |
| **Alam et al. CNN [10]** | | **5.475** |
| **Proposed** | CSIQ quality scores [17] | **5.691** |
| | LIVE quality scores [19] | **5.991** |
| | TID 2013 quality scores [18] | **5.626** |

**Table 1**: Performance comparison between different algorithms. Highlighted are the best state-of-the-art methods (handcrafted and CNN-based), as well as our results.

We compare in Table 1 the performance of our method (trained with three different IQA datasets) with other predictors of local visibility thresholds proposed in the literature. The majority of these methods use handcrafted models directly derived from psychophysical experiment, while [10] employs a convolutional neural network. The performance criterion is root-mean-squared error (RMSE) between the ground-truth and predicted thresholds. We observe that our approach provides comparable or better results to state-of-the-art methods. It should be noticed that the two methods in [10] are trained on the same dataset [9] used for test, and thus their performance represents a sort of accuracy upper bound. Conversely, the proposed method achieve competitive results even when it is trained on the TID 2013 or LIVE datasets.

A qualitative illustration of the predicted thresholds is reported in Figure 4, where we also show per picture PLCC. We observe that the perceptual thresholds produced by our approach are intuitive, e.g., thresholds are lower for relatively smooth regions (sky) and higher for more complex regions (grass, leaves). There are of course also cases in which the prediction fails (see examples in Figure 5). This mainly happens in dark patches, as discussed above, where the training datasets do not offer sufficient samples to learn robustly the visibility thresholds.

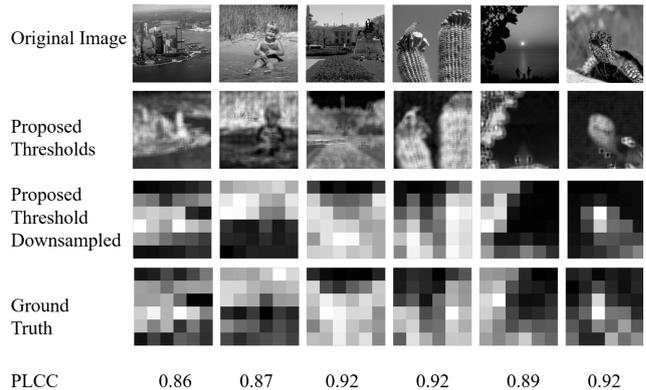

**Fig. 4**: Examples of local visibility thresholds produced by our method trained on TID 2013. The estimated thresholds are decimated to match the resolution of the ground truth. Both sets of values are scaled to fit in the range $[0, 1]$ for visualization, with 0 (black) being the lowest threshold.

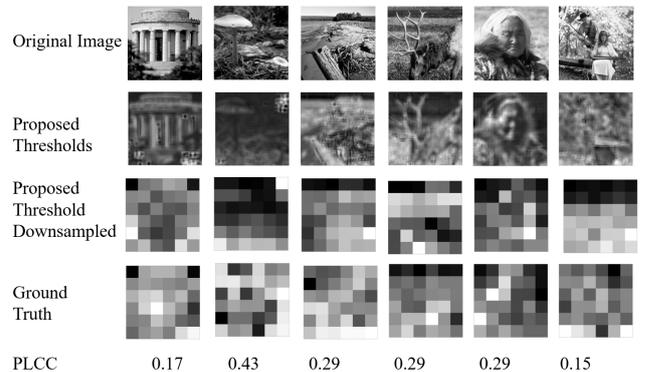

**Fig. 5**: Some failure cases of the proposed method trained on TID 2013. Often, these cases correspond to patches with low luminance, which are underrepresented in the training data.

## 4. CONCLUSION

We present a method to derive local visibility thresholds from image quality scores using a neural-network-based approach. Our experiments demonstrate that the latent information about distortion visibility carried by supra-threshold quality scores can be recovered and used to predict near-threshold local masking. One advantage of our approach, compared to models based on psychophysical data, is that it can leverage the larger availability of subjectively annotated image quality datasets. We plan to formalize further this approach in the future and apply it to objective image quality assessment.

## 5. ACKNOWLEDGMENTS

We thank Dr Anup Basu for his insightful comments and valuable feedback.